\documentclass[twocolumn,letterpaper,pre,superscriptaddress]{revtex4}
\usepackage[T1]{fontenc}
\usepackage[latin9]{inputenc}
\usepackage{amsmath}
\usepackage{graphicx}
\usepackage{amssymb}
\usepackage{hyperref}

\begin{document}

\newcommand{\MM}{$\mathcal{M}$}

\title{Manifold boundaries give ``gray-box'' approximations of complex models}

\author{Mark K.~Transtrum}
\affiliation{Department of Physics and Astronomy, Brigham Young University, Provo, Utah 84602, USA}
\email{mktranstrum@byu.edu}

\begin{abstract}
We discuss a method of parameter reduction in complex models known as the Manifold Boundary Approximation Method (MBAM).  This approach, based on a geometric interpretation of statistics, maps the model reduction problem to a geometric approximation problem.  It operates iteratively, removing one parameter at a time, by approximating a high-dimension, but thin manifold by its boundary.  Although the method makes no explicit assumption about the functional form of the model, it does require that the model manifold exhibit a hierarchy of boundaries, i.e., faces, edges, corners, hyper-corners, etc.  We empirically show that a variety of model classes have this curious feature, making them amenable to MBAM.  These model classes include models composed of elementary functions (e.g., rational functions, exponentials, and partition functions), a variety of dynamical system (e.g., chemical and biochemical kinetics, Linear Time Invariant (LTI) systems, and compartment models), network models (e.g., Bayesian networks, Markov chains, artificial neural networks, and Markov random fields), log-linear probability distributions, and models with symmetries.  We discuss how MBAM recovers many common approximation methods for each model class and discuss potential pitfalls and limitations.  
\end{abstract}

 
\maketitle

\section{Introduction}
\label{sec:introduction}

Mathematical models have always played an import role science.  They are a key way of summarizing and exchanging complex information, evolve with the acquisition of new knowledge, and motivate starting points for future explorations.   Driven in part by big data and expanding computational capabilities, models with many parameters (in some cases thousands or more) are increasingly common.  These models create new challenges and raise new questions ranging from the technical to the profound.  Large-scale models may introduce numerical instability, be computationally expensive, or be challenging to fit to data.  Methods to manage model complexity are a central concern in modern science.

Model reduction is the problem of finding approximate, simpler models that capture the same behavior as the original model.  Simple models have many motivations.  They reduce computational cost and avoid technical numerical problems.  They avoid statistical challenges associated with model calibration and over-fitting.  They are easier to interpret and so may also help reveal the important mechanisms that drive a particular behavior.  

Model reduction is a vast topic and a comprehensive review of the literature would be impossible.  Although there is not a one-size-fits all solution, there are a host of available methods that have found consistent success within their appropriate scope.  Of particular note are methods that exploit scale separation, particularly singular perturbation methods\cite{saksena1984singular,kokotovic1999singular,naidu2002singular}, lumping or clustering methods\cite{wei1969lumping,liao1988lumping,huang2005systematic}, and many others\cite{conzelmann2004reduction,antoulas2005approximation}.  Of particular note are methods developed by the control and chemical kinetics communities focused on dynamics systems\cite{moore1981principal,saksena1984singular,kokotovic1999singular,naidu2002singular,antoulas2005approximation,lee2010multi,dullerud2013course}.  In physics, techniques such as mean-field theory and the Renormalization Group (RG) are also powerful techniques\cite{goldenfeld1992lectures,zinn2007phase}.

Typically, choosing a model reduction algorithms is determined by the functional form of the model to be reduced and the motivation behind the reduction.  Many methods are automatic methods that give ``black box'' approximations that are computationally efficient surrogates for the original model.  One possible motivation for these ``black box'' approximations is to enable real-time control of a complex system in cases where evaluating the complete model would be too slow or otherwise infeasible. 

In this work, we are particularly interested in models motivated by mechanistic information about a system.  In that case, a good model reduction method preserves the relevant mechanistic information to produce a ``gray-box'' description that balances model complexity with mechanistic fidelity\cite{verghese2009getting,kourdis2010physical}.  In this case, model reduction can help to identify the relevant, governing mechanism that governs a particular behavior, i.e., distinguish between the relevant and irrelevant details.  For example, the Renormalization Group (RG) quantifies the stability of macroscopic observations to microscopic variations in structure, justifying the use of coarse-grained, simplified models.  It also makes concrete the concepts of relevant and irrelevant degrees of freedom by identifying those control knobs that must be tuned to observe a particular behavior.  These methods are not without their limitations; applications typically require that systems exhibit a high-degree of symmetry in the microscopic interactions.

Information theory has recently been invoked to justify why simple, effective models may be quantitatively predictive in many other complex systems\cite{machta2013parameter,transtrum2015perspective}.  This approach uses the Fisher Information Matrix (FIM) to quantify the relative importance of parametric degrees of freedom in a complex model.  Small eigenvalues of the FIM therefore correspond to irrelevant details that could, in principle, be discarded from the model.  This approach complements and extends traditional arguments based on renormalization group or continuum limits.  Where applicable, methods such as RG can construct simple macroscopic representations from microscopic models.  However, these techniques break down on models of complex, heterogeneous systems.  Nevertheless, these models (sometimes known as ``sloppy models'') will often have many small FIM eigenvalues\cite{brown2003statistical,brown2004statistical,waterfall2006sloppy,gutenkunst2007universally,chachra2012structural}, giving hope that simple, parsimonious representations of these systems may yet be found that transparently bridge microscopic mechanisms with macroscopic phenomenology.

A candidate reduction method that grew out of the information theory approach to modeling is the Manifold Boundary Approximation Method (MBAM)\cite{transtrum2014model}.  The foundation for this approach is a geometric interpretation of statistics in which the FIM acts a Riemannian metric on the model's parameter space\cite{rao1949distance,bates1980relative,murray1993differential,amari2007methods,transtrum2010nonlinear,transtrum2011geometry}.  The basis for this approach is the observation that sloppy models often correspond to manifolds that are bounded with a hierarchy of widths.  The MBAM identifies the boundary oriented with the principal components of the manifold and uses this boundary as an approximation.  The MBAM operates iteratively, removing one parameter a time, leading to a sequence of approximate models that connect microscopic components with systems-level behavior.  It is therefore a promising method for identifying the mechanistic causes of system-behavior.  

The MBAM is a general method that makes few assumptions about the mathematical form of the model.  In principle it can be applied to models of dynamical systems as well as field theories, for example.  However, the MBAM requires that the model have a particular \emph{global} structure: a model manifold with a hierarchical boundary structure with a sequence of faces, edges, corners, hyper-corners, etc..  Because this structure is invariant to diffeomorphisms of the model manifold, it has been described as the \emph{information topology} of the model manifold\cite{transtrum2014information}.  

This paper has two primary objectives.  First we elaborate on the details of the manifold boundary approximation method.  Second, we use computational differential geometry to explore the boundary structure of several model classes and demonstrate how MBAM would be realized on these models.  We show that the necessary hierarchical boundary structure is common to many model classes.  In many cases, boundaries have been implicitly used by insightful modelers seeking effective approximations.  In these cases, the MBAM would have identified these approximations in a semi-automatic way.  The paper is organized as follows: In the next section, we give a review of the manifold boundary approximation method, followed by a simple illustrative example.  In section~\ref{sec:Examples}, we consider several different model classes one-by-one, including models composed of elementary operations, dynamical systems, network models, log-linear distributions such as the Ising model, and models with symmetries.  We discuss functional similarities to other model reduction methods in section~\ref{sec:relation}, as well as limitations of the MBAM in section~\ref{sec:limitations}.  Finally we discuss some of the implications of MBAM in section~\ref{sec:Discussion}.

\section{The Manifold Boundary Approximation Method (MBAM)}
\label{sec:mbam}

\subsection{Algorithmic Description}
\label{sec:mbamalgorithm}

The Manifold Boundary Approximation Method (MBAM) is a model reduction scheme described in reference\cite{transtrum2014model}.  As the name suggests, it is based on a geometric interpretation of information theory (known as information geometry\cite{rao1949distance,beale1960confidence,bates1980relative,murray1993differential,amari2007methods,transtrum2010nonlinear,transtrum2011geometry}) that aims to bridge underlying mechanisms with the system-behavior in a wide range of model types.  In this section we give a more detailed presentation than what was originally described in reference\cite{transtrum2014model}.  In-depth applications to biological models are given elsewhere\cite{transtrum2015bridging}.

We assume the existence of a model in the form of a probability distribution with parameter vector $\theta$.   Since approximations such as the MBAM necessarily disregard pieces of the model, it is necessary to identify the objective in mind, i.e., which model behaviors the approximation should preserve.  We refer to the particular system behaviors that should be preserved under model reduction as Quantities of Interest (QoIs) which we denote by $\xi$.  The model $P(\xi, \theta)$ gives the probability of observing the QoIs given parameters $\theta$.  

There is no general rule for choosing QoIs.  In practice, the QoIs may often include predictions for which experimental data is available.  The data will then be used to calibrate the reduced model.  However, QoIs may also include predictions for which data is unavailable but for which the modeler would nevertheless like to make predictions.  Alternatively, QoIs may include a very small subset of possible predictions in order to identify a minimal characterization of a system behavior.  In other cases, the QoIs may be the probability of all possible predictions, for example as in statistical mechanics which we consider later.  In any case, identifying appropriate QoIs for the application in mind is an important step in applying MBAM.

The underlying idea of the MBAM is that the function $P(\xi, \theta)$ is a vector in an inner-product space.  If the model contains $N$ parameters, then this vector sweeps out an $N$-dimensional hyper-surface embedded in this space.  This hyper-surface is known as the model manifold and denoted by \MM.  For many systems, the model manifold is bounded with cross-sections forming an exponential hierarchy of widths.  Consequently, \MM \ often has an \emph{effective dimensionality} that is much less than $N$.  Our goal is to construct a low dimensional approximation to the model manifold by finding the boundaries of \MM.  The procedure for doing this can be summarized as a four step algorithm.

First, from an estimate of the parameters $\theta_0$ calculate the Fisher Information Matrix (FIM)
\begin{equation}
  \label{eq:FIM}
  g_{\mu\nu} = - \left\langle \frac{\partial^2 \log P}{\partial \theta_\mu \theta_\nu} \right\rangle = \left\langle \frac{\partial \log P}{\partial \theta_\mu}  \frac{\partial \log P}{\partial \theta_\nu}  \right\rangle.
\end{equation}
The FIM acts as a Riemannian metric on \MM.  Calculating the eigenvalues of this matrix reveals the ill-conditioned nature of the corresponding parameter inference problem.  The eigenvectors with small eigenvalues correspond to the parameter combinations that would be unidentifiable from an observation of the QoIs and so have little explanatory value.  We denote the eigenvector with smallest eigenvalue by $v_0$.  

The second step is to calculate a parameterized path through parameter space $\theta(\tau)$ corresponding to the geodesic originating with parameters $\theta_0$ and direction $v_0$.  This is found by numerically solving a differential equation:
\begin{equation}
  \label{eq:geodesic}
  \frac{d^2}{d \tau^2} \theta^\mu = \Gamma^\mu_{\alpha \beta} \frac{d \theta^\alpha}{d\tau} \frac{d \theta^\beta}{d\tau}
\end{equation}
where
\begin{equation}
  \label{eq:Gamma}
  \Gamma^\mu_{\alpha\beta} = g^{\mu\nu} \left\langle  \frac{\partial \log P}{ \partial \theta_\nu} \left(  \frac{\partial^2 \log P}{\partial \theta_\alpha \partial \theta_\beta} + \frac{1}{2} \frac{\partial \log P}{\partial \theta_\alpha}  \frac{\partial \log P}{\partial \theta_\beta}   \right) \right\rangle
\end{equation}
is the Riemann connection on the manifold and $g^{\mu\nu} = \left(g^{-1}\right)^{\mu\nu}$. 

Computational cost is always a concern for methods targeted at large models.  The most expensive part of this calculation is evaluating the FIM which requires calculating the derivative of the model with respect to all parameters and thus scales linearly with the number of parameters.  These derivative calculations are trivially parallelized.  In contrast, calculating $\Gamma^\mu_{\alpha\beta} \dot{\theta}^\alpha \dot{\theta}^\beta$ requires only an additional second directional derivative.  This can be estimated with computational cost comparable to a single evaluation of the model.  Thus, the additional overhead of calculating a geodesic beyond the FIM becomes negligible as models become large.  However, the FIM must be evaluated repeatedly while solving Eq.~\eqref{eq:geodesic}.  We discuss the implications of this in section~\ref{sec:limitations}.

The solution to Eq.~\eqref{eq:geodesic} is a parameterized curve through the parameter space.  Along this curve, the modeler monitors the eigenvalues of the FIM.  A boundary of the model manifold is identified by the smallest eigenvalue of $g_{\mu\nu}$ approaching zero.  When the smallest eigenvalue becomes much less than the next smallest, then the corresponding eigendirection reveals a limiting approximation in the model.

The approximation will typically correspond to one or more parameters approaching zero or infinity in a coordinated way.  The third step is to identify this limit and analytically evaluate it in the model.  This step often requires some theoretical insight; several examples of how to do this are given later.  The result of the process is a new model with one less parameter.  We denote this reduced model by $\tilde{P}(\xi, \phi)$, where $\phi$ are the reduced set of parameters.

Finally, the values of the parameters $\phi$ in the approximate model are calibrated to the parameters $\theta_0$ by minimizing the information distance to the original model:
\begin{equation}
  \label{eq:LSestimate}
  \min_\phi \left\langle  \log P(\xi, \theta_0) - \log \tilde{P}(\xi, \phi)  \right\rangle
\end{equation}
where $\langle \cdot \rangle$ means expectation value with respect to the original distribution $P(\xi, \theta_0)$.  Because the first term in Eq.~\eqref{eq:LSestimate} is a constant, calibrating the model is equivalent to maximizing the log-likelihood that the model $\tilde{P}(\xi, \phi)$ generated the data.

This four-step procedure is iterated, removing one parameter at a time, until the model becomes sufficiently simple.  

The procedure just described requires a few comments.  First, the MBAM requires a parameter estimate as a starting point $\theta_0$, which usually cannot be estimated accurately. The final reduced model is largely independent to these uncertainties.  The reason for this is seen by considering a geometric argument given in reference\cite{transtrum2014model}.  Huge variations in parameter values can result when fitting to data, but these variations all lie within the same statistical confidence region, which means they map to nearby points on the model manifold.  Starting from any points within this confidence region will identify the same sequence of boundaries as the true parameters.  In other words, choosing a $\theta_0$ is incidental to the procedure but unnecessary for the final result.  

Because of its geometric motivation, the reduced models are invariant to changes in a model's parameterization, such as using rates vs. time constants.  These transformations are equivalent to coordinate transforms on the manifold.  In many applications, the microscopic parameters are restricted to positive values.  In order to guarantee positivity, it is helpful to use log-transformed parameters in the model.  This serves the dual purpose of non-dimensionalizing the parameters, that is important for the initial eigendirection of the FIM to point to the narrowest width of the \MM.  

Finally, MBAM is a nonlocal approximation in the sense that the reduced model approximates not just the behavior at $\theta_0$ but at all other nearby behaviors, i.e., those behaviors along the long-axis of the manifold.  In general, identifying all model behaviors using a brute-force exploration of parameter space is infeasible for models with more than a handful of parameter.  MBAM is able to approximate large regions of parameter space without a direct exploration by exploiting the hierarchical boundary structure of many models.  By choosing a boundary oriented with the principal axis of the manifold, the MBAM avoids the need to explicitly explore the entire parameter space.

\subsection{Illustrative Example: Biological Adaptation and Negative Feedback}
\label{sec:example}

\begin{figure}
  \includegraphics[width=3.5in]{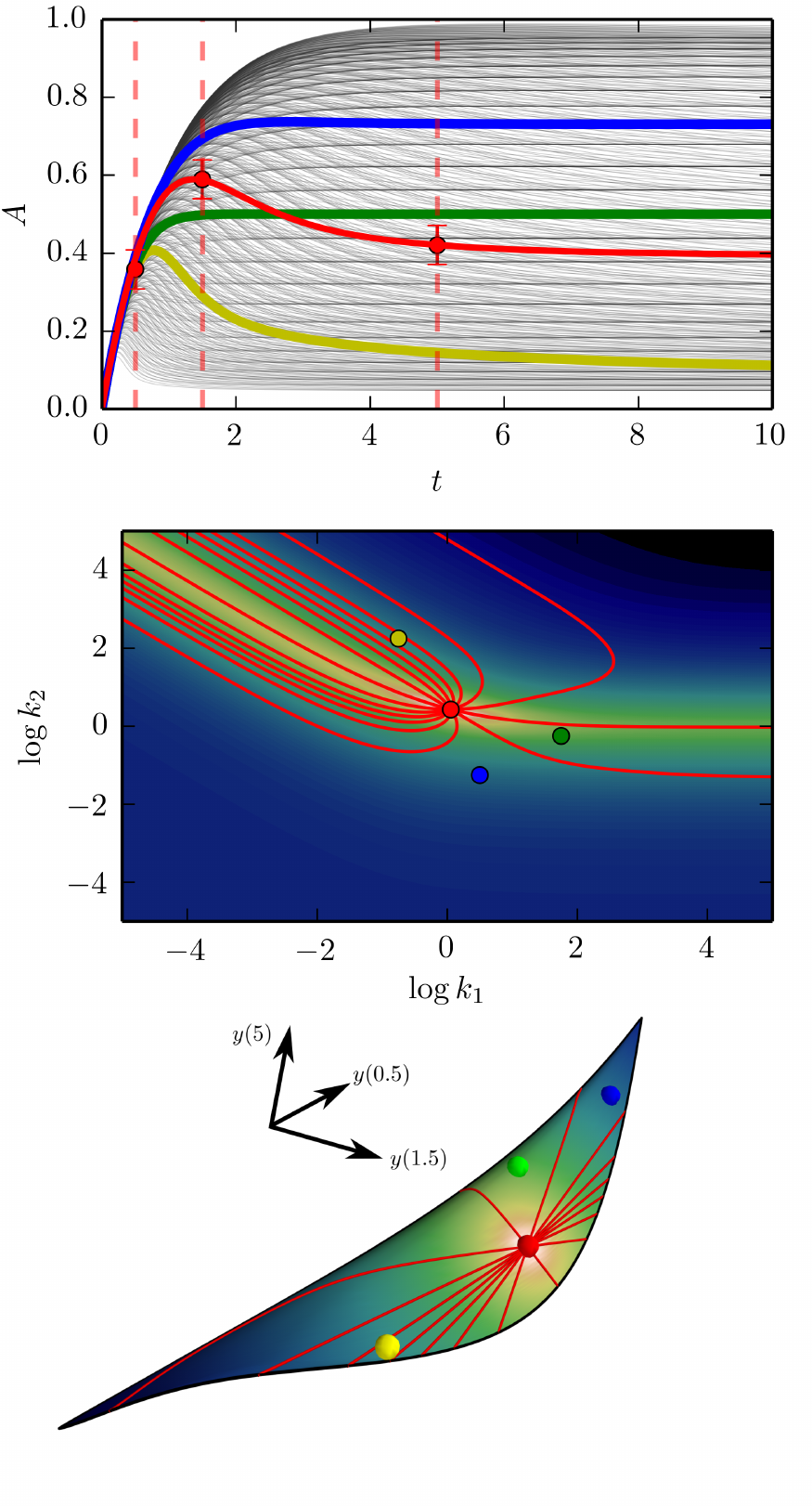}
  \caption{\label{fig:NFBLBMM} \textbf{Visualizing ranges of model behavior.}  Top: Varying the parameters of the model leads to a family of time series.  Center: Fitting these time series to data (red dots in top panel) generates a chi-squared cost surface in parameter space.  Bottom: The Model Manifold \MM \ is generated by collecting all possible predictions of the model at specific time points (dashed, red vertical lines in top panel).  Geodesics are the analogs of straight lines on curved surfaces and are depicted by the red curves in the center and bottom panels.}
\end{figure}

We illustrate the manifold boundary approximation method with a simple example.  Consider a simple two-parameter model that arises in the study of biological adaptation to the mean\cite{sontag2008remarks,ma2009defining,nemenman20124,transtrum2015bridging}:
\begin{eqnarray}
  \label{eq:nfblb1}
  \frac{dA}{dt} & = & (1 - A) - k_2 A B \\
  \label{eq:nfblb2}
  \frac{dB}{dt} & = & k_1 A (1 - B)
\end{eqnarray}
with $k_1$ and $k_2$ as parameters and $A$ and $B$ dynamical variables with initial conditions zero.  We take as quantities of interest the time series for $A$ with additive Gaussian noise leading to a least-squares estimate when fit to data.  Varying parameter values leads to different time series for this model as in Figure~\ref{fig:NFBLBMM} (top).   Fitting this model to a single realization of the data (red dots in Figure~\ref{fig:NFBLBMM}, top) leads to a chi-squared cost surface (Figure~\ref{fig:NFBLBMM}, center).  By considering all possible model predictions for the QoIs in data space, we generate the model manifold (Figure~\ref{fig:NFBLBMM}, bottom).

The red curves in Figure~\ref{fig:NFBLBMM} center and bottom panels are geodesics on \MM \ represented in parameter space and behavior space respectively.  Notice how the geodesics locally align with the cost surface while globally exploring the model manifold.  In this way geodesics connect local information, i.e., the Fisher Information, with the model's nonlocal structure, i.e. boundaries.  

Manifold boundaries correspond to physically realistic models; the boundaries oriented with the principal axes of the manifold are natural approximations to the full model.  Consider the geodesics moving toward the upper-left corner of Figure~\ref{fig:NFBLBMM}, middle panel.  These curves ultimately terminate at the lower boundary of the manifold in Figure~\ref{fig:NFBLBMM}, bottom panel.  This boundary corresponds to the limit that $k_1 \rightarrow 0$ while $k_2 \rightarrow \infty$.  Evaluating this limit in the model requires some theoretical work.  Inspecting Eqs.~\eqref{eq:nfblb1}-\eqref{eq:nfblb2}, we see that if $k_1 \rightarrow 0$, then $B \rightarrow 0$.  However, the influence of $B$ on $A$ is scaled by the parameter $k_2$ which becomes very large.  Therefore, we define a \emph{renormalized} variable $\tilde{B} = k_2 B$, and the model becomes
\begin{eqnarray}
  \label{eq:nfblbrenorm1}
  \frac{dA}{dt} & = & (1 - A) - A \tilde{B} \\
  \frac{d \tilde{B}}{dt} & = & (k_1 k_2) A ( 1 - \tilde{B}/k_2) \nonumber \\
  \label{eq:nfblbrenorm2}
  & \rightarrow & \tilde{k} A,
\end{eqnarray}
where we have used the renormalized parameter $\tilde{k} = k_1 k_2$.

Eqs.~(\ref{eq:nfblbrenorm1}-\eqref{eq:nfblbrenorm2}) have a simple physical interpretation.  The model behavior does not depend on absolute scale of $B$, but on $k_2 B$.  Furthermore, the curves in Figure~\ref{fig:NFBLBMM} (top) are determined primarily by the product $k_1 k_2$ and not either parameter individually.    Consequently, simultaneously estimating $k_1$ and $k_2$ from data would have led to large uncertainties (Figure~\ref{fig:NFBLBMM}, middle panel)).  On the other hand, the parameter uncertainties in the reduced model would be much smaller.  The renormalized expressions for $\tilde{B}$ and $\tilde{k}$ therefore characterize the family of physical systems with equivalent behavior.  This ``gray-box'' representation preserves the negative feedback mechanism while discarding the information about the scale of $B$.

The aspect ratio of the model manifold in Figure~\ref{fig:NFBLBMM} is not particularly dramatic so the approximation in Eqs.~\eqref{eq:nfblbrenorm1}-\eqref{eq:nfblbrenorm2} is not very accurate for some values of the parameters.  For sloppy models with more than a few parameters, aspect ratios greater than 1000:1 are not uncommon, in which case the resulting approximate models are very accurate.  The width of the model manifold is a measure of the model error that is estimated automatically from the MBAM algorithm during the calibration step (Eq.~\eqref{eq:LSestimate}).

There is an inherent ambiguity in the MBAM because of the two possible boundaries in Figure~\ref{fig:NFBLBMM}.  Alternatively, we could have chosen the opposite boundary.  In this case, the geodesics in Figure~\ref{fig:NFBLBMM} indicate that $k_1 \rightarrow \infty$.  In this limit we have $B\rightarrow 1$ and the approximate model becomes
\begin{eqnarray}
  \label{eq:nfblbrenormalt1}
  \frac{dA}{dt} & = & (1 - A) - k_2 A \\
  \label{eq:nfblbrenormalt2}
  B & = & 1.
\end{eqnarray}
Notice how this limit is dual to the first.  In both cases, the physical unconstrained quantity is in the magnitude of $B(t)$.  In the first case we employed the approximation $B\rightarrow 0$, while in the second we employ the approximation $B \rightarrow 1$.  This illustrates a general principle: reduced models are not unique.  In general, any model can be approximated by any other model in the same universality class.  Manifold boundaries are convenient choices because structural simplifications occur at extreme values of the parameters.

When MBAM is applied to a model with many parameters, the four-step algorithm is iterated several times.  At each step there may be several approximations to choose from.  In our experience, the final model is largely independent of these choices.  To understand why, image that Figure~\ref{fig:NFBLBMM} corresponds to a two-dimensional cross section of a high-dimensional model manifold.  If the limit $k_1 \rightarrow \infty$ is selected first, then the second iteration is likely to identify the dual limit so that the approximate model corresponds to one of the two hyper-corners in Figure~\ref{fig:NFBLBMM}.  That same hyper-corner would have been reached if the order of limits had been reversed.  Ultimately, the only ambiguity is the order in which the boundaries are evaluated.

\section{Examples of Models with a Hierarchical Boundary Structure}
\label{sec:Examples}

Having illustrated the basic process of the MBAM, we now begin to enumerate a catalog of model classes that have a hierarchical boundary structure, i.e., faces, edges, corners, hyper-corners, etc.  This list is not complete, and some counterexamples will be given section~\ref{sec:limitations}.

\subsection{Compositions of Elementary Functions}
\label{sec:compositions}

\begin{figure}
  \includegraphics[width=3.5in]{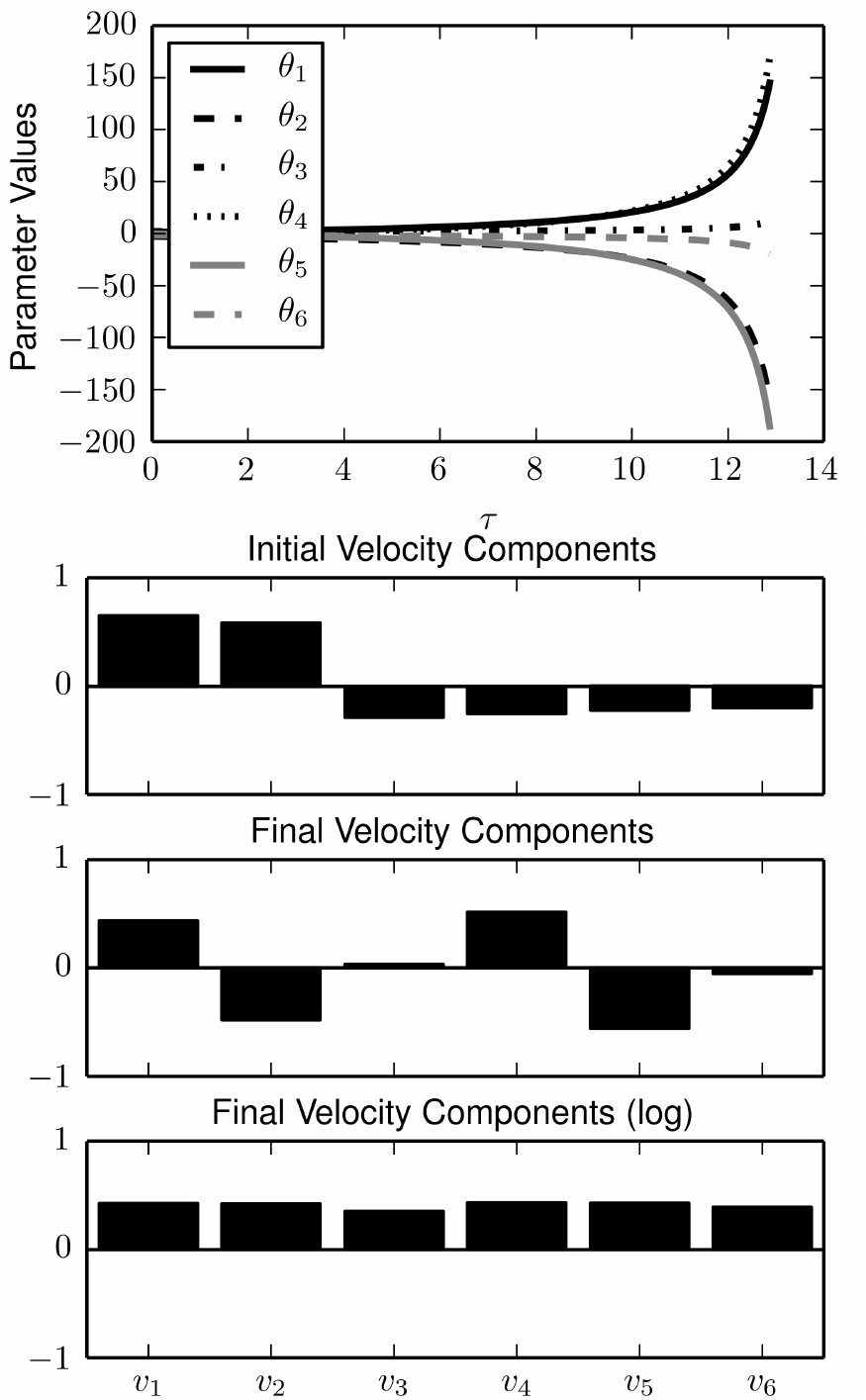}
\caption{\label{fig:RationalGeodesic} \textbf{Geodesics for a rational function.}  Top: Parameter values along the geodesic for the model in Eq.~\eqref{eq:rational}.  Notice the geodesic encounters a singularity around $\tau = 13$.  The components of the geodesic velocity are shown in the next three panels.  First the initial components of the velocity are determined by the eigenvector of the FIM with smallest eigenvalue.  Next, the final components of the geodesic velocity near the singularity at $\tau = 13$.  These components satisfy the equation $d\phi = 0$ as explained in the text.  This is made clear by considering the velocity components in the log-parameters (bottom panel), in which case the components are equal.}
\end{figure}

\textbf{Rational Functions:}
We first consider several functions composed from elementary operations.  For example, consider the six parameter rational function
\begin{equation}
  \label{eq:rational}
  y(t, \theta) = \frac{t^2 + \theta_1 t  + \theta_2}{\theta_3 t^3 + \theta_4 t^2 + \theta_5 t + \theta_6}.
\end{equation}
We take as QoIs the output of this model at several time points with additive Gaussian noise.  The geodesic for this model is summarized in Figure~\ref{fig:RationalGeodesic}.  The manifold boundary is identified by a singularity that occurs around $\tau = 13$.  At this point, the absolute value of several of the parameters are growing without bound.  

In this case, knowing that the parameters become infinite (in absolute value) at the boundary is sufficient to identify the reduced model.  Dividing the numerator and denominator by $\theta_6$, for example, gives
\begin{eqnarray}
  \tilde{y}(t, \theta) & = & \frac{t^2/\theta_6 + (\theta_1/\theta_6) t  + (\theta_2/\theta_6)}{(\theta_3/\theta_6) t^3 + (\theta_4/\theta_6) t^2 + (\theta_5/\theta_6) t + 1} \nonumber \\
  \label{eq:rationalapprox}
  & \rightarrow & \frac{ \phi_1 t  + \phi_2}{\phi_3 t^3 + \phi_4 t^2 + \phi_5 t + 1} \nonumber \\
\end{eqnarray}
where the renormalized parameters are $\phi_\mu = \theta_\mu/\theta_6$.

Consider the hyper-surfaces in parameter space defined by $\phi_\mu$ = constant.  These hyper-surfaces are very nearly parallel to the manifold boundary so that the unconstrained parameter combination must be perpendicular to this surface.  Normals to these surfaces given by the differential form $d \phi_\mu = 0$.  Since the geodesic velocity is approximately parallel to the unconstrained parameter combination, the components of the geodesic velocity are often a useful clue for identifying the boundary approximation and the reduced parameters as we now illustrate.

In the second and third panels of Figure~\ref{fig:RationalGeodesic} we give the initial and final) components of the geodesic velocity.  In this case, the components of the final velocity all have different magnitudes.  These magnitudes are determined by the requirement $d \phi_\mu = 0$ which gives $d \theta_\mu = (\theta_\mu / \theta_6) d \theta_6$.  We confirm that the relative height of the peaks in Figure~\ref{fig:RationalGeodesic} middle panel are given by this relation.  Indeed, considering log-transformed parameters, we have $d \phi_\mu = 0$ implies that $ d \log \vert \theta_\mu \vert = d \log \vert \theta_6 \vert$ as shown in Figure~\ref{fig:RationalGeodesic} bottom panel.  This example demonstrates that the boundaries and corresponding limits identified by MBAM are invariant to reparameterization of the model.

By iterating the MBAM, we remove additional parameters.  In the next iteration, $\phi_\mu \rightarrow \infty$, the model becomes:
\begin{equation}
  \label{eq:rational2approx}
  \tilde{\tilde{y}}(t) = \frac{ (\phi_1/\phi_5) t  + (\phi_2/\phi_5)}{(\phi_3/\phi_5) t^3 + (\phi_4/\phi_5) t^2 + t}.
\end{equation}

\textbf{Sums of Exponentials:}
The model corresponding to
\begin{equation}
  \label{eq:exponentials}
  y(t, A, \lambda) = \sum_\mu A_\mu e^{-\theta_\mu t}
\end{equation}
with parameters $\theta = (A_\mu, \lambda_\mu) > 0$ was considered in reference\cite{transtrum2014model} in which case it was shown that the manifold boundary approximation was equivalent to reducing the number of terms in the sum and adding a constant term.  This model is interesting, however, because the model manifold has unbounded directions.  In particular, any of the linear parameters $A_\mu$ can be taken to infinity and the model predictions will similarly become infinite.  Although the manifold is not bounded, it nevertheless has bounded cross sections that form a hierarchical structure.  We discuss the possibility of unbounded manifolds in section \ref{sec:limitations}.

\textbf{Composing Rational Functions and Exponentials:}
We now elaborate on previous examples by composing rational and exponential functions:
\begin{equation}
  \label{eq:partitionfunction}
  y(T, g, \Delta E) = \frac{g e^{-\Delta E/k_B T}}{1 + g e^{-\Delta E/k_B T}}.
\end{equation}
This function arises in modeling the probability of observing a two-state system in a particular state in thermal equilibrium as a function of temperature\cite{hansen2015enzyme}.  The parameters are $\theta = (g, \Delta E)$ where $g\geq 0$, $\Delta E$ is unbounded, and temperature $T$ is the independent variable.  

This model has a few obvious limits.  If $g \rightarrow \infty$ or if $\Delta E \rightarrow -\infty$ then $y(T) = 1$ and if $g \rightarrow 0$ or $\Delta E \rightarrow \infty$ then $y(T) = 0$.  However, neither of these are boundaries of the model manifold.  The reason is that these obvious limits remove both parameters.  Except in unusual circumstances (as we discuss in section \ref{sec:limitations}), the edge of the two-parameter model is a one-parameter model.

\begin{figure}
  \includegraphics[width=\linewidth]{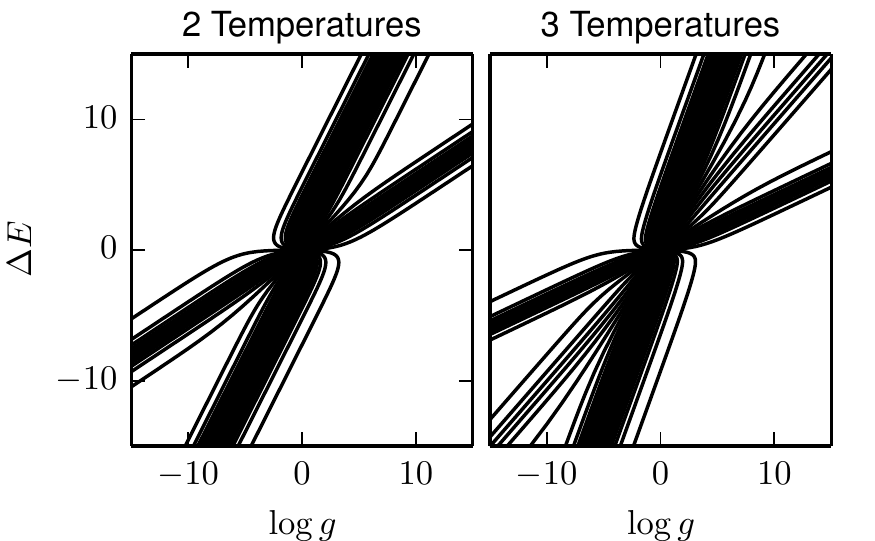}
  \caption{\label{fig:partition} \textbf{Geodesics for model in Eq.~\eqref{eq:partitionfunction}}  The different slopes at which the geodesic approaches infinity correspond to different boundaries.  Selecting two observation temperatures as QoIs leads to four boundaries (left) while three temperatures leads to six boundaries (right). }
\end{figure}

We consider the simple case in which the QoIs consist of $y(T)$ at two different temperatures.  In Figure~\ref{fig:partition} (left) we show the geodesics for this model originating from the origin in many different directions.  In this case the model has four boundaries as can be identified by the four different slopes with which the geodesic path takes parameters to infinity.

If $g\rightarrow \infty$ and $\Delta E \rightarrow \infty$, then the critical expression $\log g - \Delta E / k_B T$ will become either positive or negative infinity depending on the temperature.  Therefore, there is a critical temperature, $T_c$ at which a transition occurs: $y(T < T_c) = 0$ and $y(T > T_c) = 1$.  The reduced parameter $\phi = \log g - \Delta E / k_B T_C$ controls the value of $y(T_c)$.  The two boundaries for which $g\rightarrow \infty$ and $\Delta E \rightarrow \infty$ therefore correspond to the two possible choices of critical temperatures.  Indeed, the slopes of these geodesics paths are given by $d \phi = 0$ which gives $d \log g/ d\Delta E = 1/k_B T$.

The two boundaries for which $g\rightarrow -\infty$ and $\Delta E \rightarrow -\infty$ similarly correspond to the two possible choices of $T_c$, but with a reversed transitions: $y(T < T_c) = 1$ and $y(T > T_c) = 0$.

This analysis suggests that changing the QoIs to include $3$ temperatures would result in a model manifold with six faces, as we confirm numerically in Figure~\ref{fig:partition} (right).

In terms of the probability distribution of a two-state system, these limits have natural interpretations.  In general, there will be a transition from low to high probability with temperature.  The temperature over which this transition occurs depends on the balance between the relative multiplicities of the two states and the energy difference.  Each of the boundaries above correspond to the approximation that this transition is abrupt.  

\subsection{Dynamical Systems}
\label{sec:dynamicalsystems}

Many systems of physical interest are described by dynamical systems.  Depending on the functional form of the dynamical system, there are many classes to consider.  In the interest of space, several of these classes have been or will be discussed in more detail elsewhere, including network models of biochemical kinetics (including mass-action and Michaelis-Menten dynamics)\cite{transtrum2015bridging}, linear-time invariant systems (such as arise in control theory)\cite{pare2015unified}, models of transient dynamics in power systems, and neuroscience models of Hodgkin-Huxley neurons.  For completeness, we here note that models in each of these classes also exhibit a hierarchical boundary structure. 

Here we consider an example of a multi-compartment model that will serve as a segue into a discussion about network models in section~\ref{sec:networks}.  A compartment model describes the flow of material or energy among compartments and is common in a variety of fields including epidemiology (such as the SIR model),  pharmokinetics for describing drug delivery, ecology, and many others.  Consider here for example a three-compartment model with compartments $A$, $B$, and $C$ connected in series so that material begins in compartment $A$ and then flows to compartment $B$ and then from $B$ to $C$: $A \rightarrow B \rightarrow C$.  Assuming linear kinetics, the corresponding differential equations are
\begin{eqnarray}
  \label{eq:compartmentA}
  \frac{dA}{dt} & = & -k_1 A \\
  \label{eq:compartmentB}
  \frac{dB}{dt} & = & k_1 A - k_2 B \\
  \label{eq:compartmentC}
  \frac{dC}{dt} & = & k_2 B.
\end{eqnarray}
We take as QoIs, the values of $C$ at several times.  

\begin{figure}
  \includegraphics[width=3.5in]{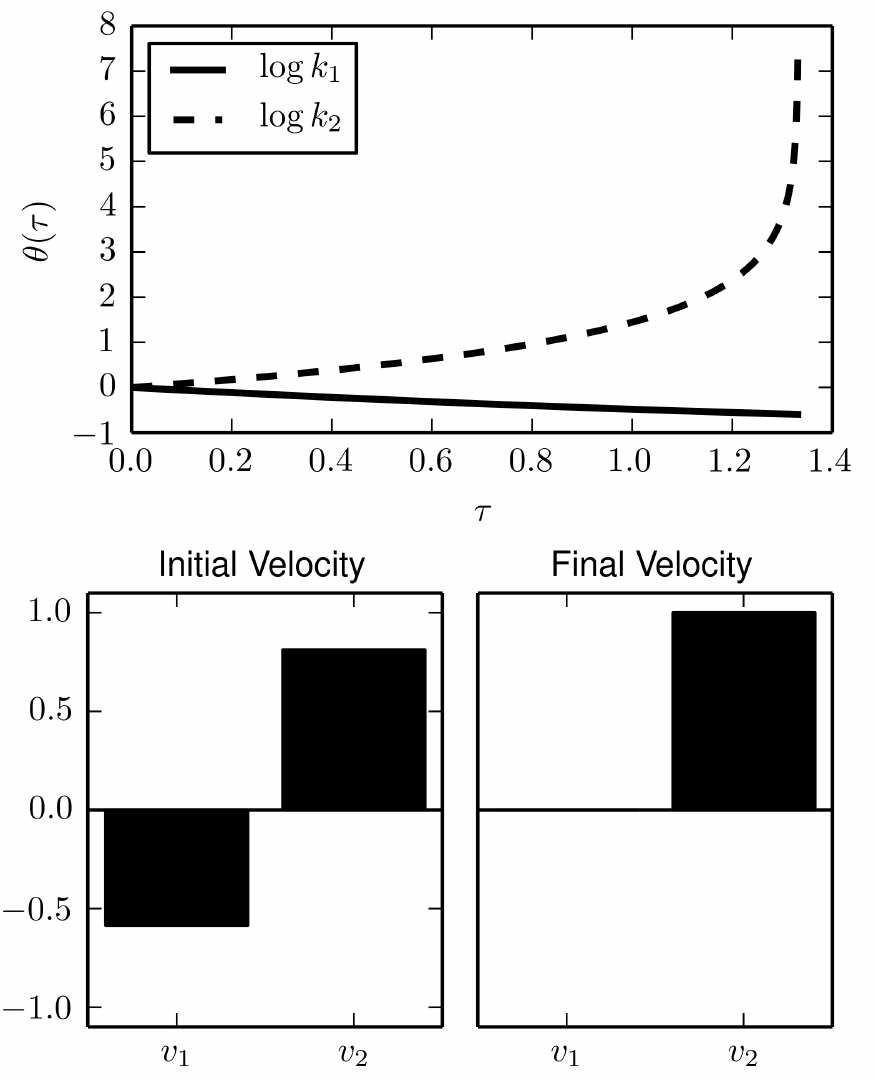}
\caption{\label{fig:ssaGeodesic} \textbf{Geodesic reveals a steady state approximation}  The manifold boundary for the model in Eqs.~\eqref{eq:compartmentA}~\eqref{eq:compartmentC} is identified by the singularity in the geodesic near $\tau = 1.3$ and corresponds to the limit that $k_2\rightarrow\infty$.  In this limit, $dB/dt = 0$, and is equivalent to a steady state approximation.}
\end{figure}

Geodesics reveal that one boundary of this model manifold is the limit $k_2 \rightarrow \infty$ (Figure~\ref{fig:ssaGeodesic}).  This limit can be evaluated by noting that $B \rightarrow 0$ so that $k_2 B = k_1 A$:
\begin{eqnarray}
  \label{eq:compartmentapproxA}
  \frac{dA}{dt} & = & -k_1 A \\
  \label{eq:compartmentapproxC}
  \frac{dC}{dt} & = & k_1 A.
\end{eqnarray}

The steady state approximation is frequently used to reduce the order of many dynamical systems, such as the Michaelis-Menten approximation which is common in biochemical kinetics.  In this case, it would typically be applied as the condition $dB/dt = 0$ which leads to the relation $k_1 A = k_2B$.  The steady state approximation is therefore mathematically equivalent to the manifold boundary approximation.  

Many other methods for deriving reduced mechanisms operate in a similar fashion.  Just as the steady-state assumptions force the rate of production of some species to vanish, partial equilibrium assumptions similarly force other rates to extreme values and therefore decouple the relevant mechanisms\cite{maas1992simplifying}.  The key difference between the is that the manifold boundary can be identified in a semi-automatic fashion without the need to have deep insights into the mechanisms driving complex systems.

In control theory, the controllability and observability of a system is quantified by the Hankel singular value\cite{dullerud2013course}.  A common approximation is to disregard the states that are least-controllable and observable, known as balanced truncation\cite{moore1981principal}.  If the Hankel singular values are treated as parameters, then balanced truncation is equivalent to the manifold boundary in which the smallest Hankel singular values become zero, i.e., the state becomes completely unobservable and controllable\cite{pare2015unified}.  

\subsection{Networks Models}
\label{sec:networks}

Many models have an associated network structure.  The problem of structure-preserving model reduction is an important one with extensive treatment in the controls community\cite{lall2003structure,sandberg2009model,yeung2011mathematical,trnka2013structured,ishizaki2014model,abad2014graph}.  In these cases it is often desirable to approximate the model that retains some semblance of the original network.  Within the context of the MBAM, if the parameters are directly related to the network structure, representing the strength of edge connections for example, then the resulting approximations have an approximate network structure directly linked to the original.

We have already seen this in the previous section for the case of a simple compartment model.  The original network ($A \rightarrow B \rightarrow C$) is automatically condensed into $A \rightarrow C$ in which the indirect flow of material from $A$ to $C$ is replaced with an direct, effective link.  Approximations like this are also important in fields for which meaning is attached to the network topology.  Systems biology, for example, is particular concerned with the appearance of network motifs\cite{shen2002network,milo2002network,mangan2003structure,alon2007network}.  In reference\cite{transtrum2015bridging} it was shown how repeated iteration of MBAM can condense a complex topology of protein reactions into a simple effective topology among clusters of proteins.

We now consider other types of network models that similarly have a hierarchical boundary structure such that iterating MBAM preserves an effective network structure.  

\textbf{Bayesian Networks and Markov Chains:} As a first example, consider the case of a Bayesian Network model.  Bayesian networks are probabilistic models formulated as directed acyclic graphs representing the conditional dependence of random variables.  Because the parameters correspond to conditional probabilities, the model manifold inherits a hierarchical boundary structure from the probabilistic meaning of the parameters.  Each parameter is positive and each set of conditional probabilities must sum to one.  The parameter space, is therefore restricted to some high-dimensional simplex. The model manifold inherits the same hierarchical boundary structure from the region of allowed parameter values.

The arguments above generalize for network structures in which parameters are themselves probabilities.  For example, Markov chains are often depicted as networks with nodes as states and edges representing transition probabilities.  These probabilities are subject to similar constraints as those in Bayesian networks leading to a hierarchical boundary structure.

\textbf{Artificial Neural Networks:} 
Another example is an artificial neural network, a common model for machine learning.  Nodes represent artificial neurons, which are typically understood to represent a sigmoidal activation function of its inputs $\sigma(I)$.  (There are many possible activation functions.  We use $\sigma(I) = \tanh(I)$, but the following results are independent of this choice.)  The network structure indicates which artificial neurons serve as inputs to other neurons.  Parameters are weights associated with edges that indicate both the strength and type (promoting or inhibitory) of the connection.

It is instructive to first consider a single neural activated by two environmental inputs
\begin{equation}
  \label{eq:ann}
  y(x_1, x_2, \theta) = \tanh(\theta_1 x_1 + \theta_2 x_2 + \theta_3)
\end{equation}
where $x_1$ and $x_2$ are the environmental inputs that activate the neuron with weight $\theta_1$ and $\theta_2$ respectively.  The third parameter $\theta_3$ is a bias term.  

The boundaries of this model correspond to the limit in which $\vert \theta_\mu \vert \rightarrow \infty$ for $\mu$ = 1, 2, and 3 with reduced parameters $\phi_1 = \theta_2/\theta_1$ and $\phi_2 = \theta_3/\theta_1$.  In this limit the output of the neuron is $\pm$1 for any values of $x_1$ and $x_2$.  The reduced model is therefore
\begin{equation}
  \label{eq:annred}
  \tilde{y}(x_1, x_2, \phi) = \begin{cases}
1 & \mbox{if } x_1 > -\phi_1 x_2 - \phi_2 \\
-1 & \mbox{if } x_2 < -\phi_1 x_2 - \phi_2.
\end{cases}
\end{equation}
The two parameters therefore define a line through the input plane that partition inputs based on how it is classified by the reduced model.

This reduced model is equivalent to another machine learning model known as a perceptron.  Perceptrons are binary, linear classifiers with a long history going back nearly half a century\cite{rosenblatt1958perceptron}.  Although the relationship between artificial neurons (as in Eq.~\eqref{eq:ann}) and perceptrons is well-known (networks of these neural often known as multi-layer perceptrons), of particular note in this context is that this equivalence is naturally recovered by interpreting the perceptron as a boundary approximation to an artificial neuron. 

Combining several artificial neurons into a network produces a richer boundary structure in the model.  However, these boundaries can be understood as a generalization of the single-neuron boundary just considered.  Several iterations of the MBAM effectively lead to a composition of several perceptrons.  The resulting model is a binary classifier that can approximate more complex divisions of the input space, i.e., not just a single line.  The resulting calculation is closely related to yet another machine learning algorithm: Support Vector Machines (SVMs).  The relationship among SVMs, perceptrons, and artificial neural networks is also known\cite{freund1999large}.  In the current context, the interesting result is that this relationships is naturally captured in the differential topological structure of the artificial neural network model and automatically recovered by iterative application of the MBAM.

\textbf{Markov Random Fields:}  
The final network model class that we consider here is the Markov Random Field (MRF).  Like Bayesian networks, MRFs aim to represent probabilistic dependence among random variables.  While Bayesian networks are acyclic, MRFs may be cyclic.  While the hierarchical boundary structure of Bayesian networks followed naturally from the probabilistic interpretation of the parameters, the ranges of allowed parameters of a MRF may be unbounded.  The boundary structure of the model manifold in Markov random fields has a different explanation.

\begin{figure}
\includegraphics[width=3.5in]{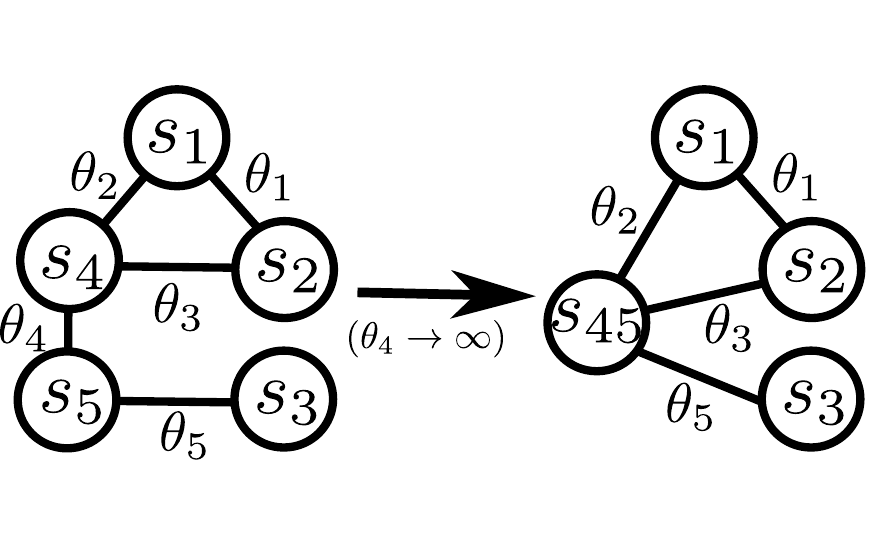}
\caption{\label{fig:mrfexmaple} \textbf{MBAM on a Markov Random Field} Left: Network structure for the Markov Random Field in Eq.~\eqref{eq:MRFPrecisionMatrix}.  Right: The manifold boundary corresponding to $\theta_4\rightarrow\infty$ corresponds to a clustering of nodes 4 and 5.}
\end{figure}

For concreteness, consider the network in Figure~\ref{fig:mrfexmaple} (left).  Each node is a random variable that can take values $\pm$ 1.  The probability of a particular configuration is the proportional to
\begin{equation}
  \label{eq:mrfprob}
  P(\mathbf{s}, \theta) \propto \exp \left( -\frac{1}{2} \, \mathbf{s}^T A(\theta) \mathbf{s} \right)
\end{equation}
where $\mathbf{s} = (s_1, s_2, s_3, s_4, s_5)^T$ and
\begin{equation}
  \label{eq:MRFPrecisionMatrix}
  A(\theta) = \left( \begin{array}{ccccc} 0 & \theta_1 & 0 & \theta_2 & 0 \\
               \theta_1 & 0 & 0 & \theta_3 & 0  \\
               0 & 0 & 0 & 0 & \theta_5 \\
               \theta_2 & \theta_3 & 0 & 0 & \theta_4 \\
               0 & 0 & \theta_5 & \theta_4 & 0 
\end{array} \right).
\end{equation}
Notice that the nonzero entries of $A$ reflect the network structure in Figure~\ref{fig:mrfexmaple}.  There is a missing normalization in Eq.~\eqref{eq:mrfprob} that depends on the parameters $\theta$ but not on the state vector $\mathbf{s}$.

For this model the manifold boundaries correspond to the limits $\theta_\mu \rightarrow \pm \infty$.  These limits correspond to a type of ``clustering'' of network nodes.  For example, if $\theta_4 \rightarrow \infty$, then the random variables $s_4$ and $s_5$ become perfectly correlated.  This reduced model therefore corresponds to the condensed network given in Figure~\ref{fig:mrfexmaple} (right).

Interesting, is the random variables can be any real number, then the model no longer has a boundary structure.  We discuss this further in section~\ref{sec:limitations}.

\subsection{Log-Linear Discrete Distributions}
\label{sec:loglinear}

We now consider probabilistic models with a finite number of outcomes such that the probability of outcome $i$ may be written as
\begin{equation}
  \label{eq:loglinear}
  P_i \propto \exp \left(  \sum_\mu \Pi_{i\mu} \theta_\mu \right).
\end{equation}
The proportionality constant depends on the parameters but is the same for each outcome.  Apart from an over-all normalization, models of this form are log-linear in the parameters.  The Markov random field from the previous section falls into this category since the matrix $A(\theta)$ is linear in the parameters.  Many standard statistical mechanics models falls into this category including the Ising model and generalizations such as the Potts model and cluster expansions of alloy formation enthalpies.  

Models of this class can be shown to always have a hierarchical boundary structure that is closely related to the structure of the model's $\Pi$ matrix.  In particular, we interpret each row of the $\Pi$ matrix as a vector in parameter space.  The convex hull of this set of points will generally have a hierarchical structure that is diffeomorphic to, i.e., has the same boundary structure as, the model manifold.

To understand this correspondence, we first assume that the model has $N$ parameters and the points in $\Pi$ are not constrained to a linear subspace of the parameter space.  This requirement means that the convex hull of points in $\Pi$ has non-zero volume.  In this case the FIM for the model is not singular at any finite values of the parameters, and any manifold boundaries must correspond to infinite parameter values.  Note that this requirement guarantees that $\Pi$ has $N$ non-zero singular values, although the converse is not true.

We now consider the behavior of the model at infinite parameter values.  These limits are characterized by some outcomes occurring with zero probability in the model (effectively freezing out the highest-energy configurations).  For any parameter values, the relative probability of observing two states is related to the difference in their $\Pi$ matrix rows:
\begin{equation}
  \label{eq:relativeprob}
  \frac{P_i}{P_j}= \exp \left(  \sum_\mu \left( \Pi_{i\mu} - \Pi_{j\mu} \right) \theta_\mu \right).
\end{equation}

Now consider a parameterized path through parameter space $\theta(\tau)$.  We assume that for large $\tau$ the path moves infinitely far from the origin and approaches a straight line, so that near the geodesic singularity we may write $\theta(\tau) = \theta_0 + v \tau$ for some vector $v$ that becomes very large but does not rotate.  In this case, Eq.~\eqref{eq:relativeprob} becomes
\begin{equation}
  \label{eq:relatieprob2}
   \frac{P_i}{P_j} = e^{\Delta \Pi \theta_0} e^{\Delta \Pi v \tau}
\end{equation}
where $\Delta \Pi = \Pi_{i\mu} - \Pi_{j\mu}$.

\begin{figure}
\includegraphics[width=3.5in]{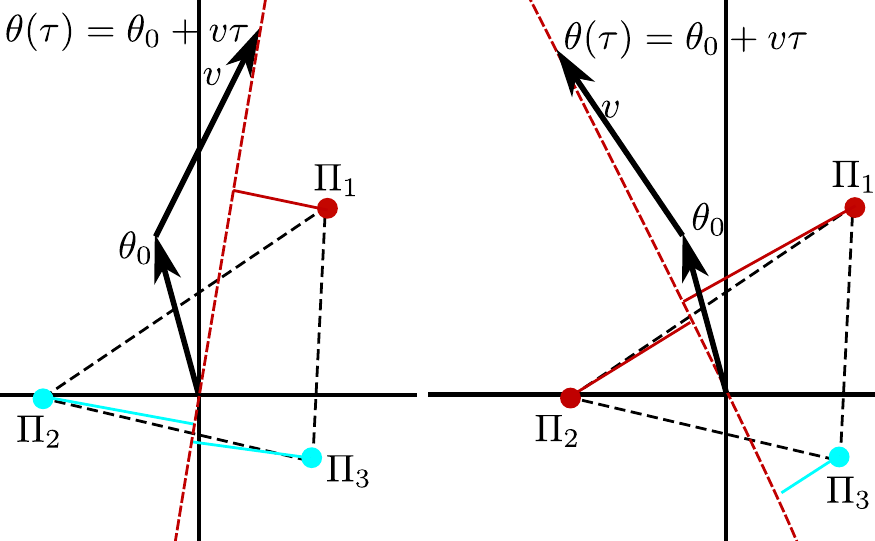}
\caption{\label{fig:convexhull} \textbf{Convex Hull of $\Pi$ determines the boundary structure.}  Rows of $\Pi$ in Eq.~\eqref{eq:loglinear} represent discrete outcomes in the model and correspond to points in parameter space (dark red and light blue dots).  Here we choose a model with three outcomes. The two parameters control the relative probabilities $P_1/P_2$ and $P_1/P_3$.  \textbf{Left}: Consider the path through parameter space given by $\theta(\tau) = \theta_0 + v \tau$ as shown.  The relative probability of each outcome is determined by the the projection of the corresponding point onto the line connecting $\theta$ to the origin (the solid dark red and light blue lines indicate this projection onto the dashed red line).  In the limit $v \rightarrow \infty$, only the outcome corresponding to $\Pi_1$ (the dark red dot) occurs with nonzero probability (the ground state).  Therefore, in this limit the model becomes insensitive to all parameters and corresponds to a vertex (i.e., zero-dimensional boundary) of the model manifold.  \textbf{Right}:  Consider the direction depicted on the right that is perpendicular to the face of the convex hull.  In the limit $v \rightarrow \infty$ the difference in the projections of $\Pi_1$ and $\Pi_2$ onto the line from $\theta$ to the origin becomes the same and both outcomes occur with nonzero probability.  In this limit, the model remains sensitive to one parameter combination that controls the relative probability of these two outcomes while the third outcome is frozen out.  This limit corresponds to an edge (i.e., one-dimensional boundary) of the model manifold.  The model manifold has the same boundary structure as that of the convex hull of $\Pi$, in this case a triangle (represented by the black dashed line).}
\end{figure}

As $\vert v \vert \rightarrow \infty$, Eq.~\eqref{eq:relatieprob2} suggests that $P_i/P_j$ will become either $0$ or $\infty$ depending on the sign of $\Delta \Pi v$ and corresponding to the cases $P_i \rightarrow 0$ or $P_j \rightarrow 0$ respectively.  For $P_i$ to remain nonzero in the limit $\vert v \vert \rightarrow \infty$, the point $\Pi_i$ must have the largest projection onto the vector $v$ of any other points in the $\Pi$ matrix.  Since $\vert v \vert \rightarrow \infty$ is the limit of zero temperature, the outcomes for which $P_i \neq 0$ correspond (typically) to ground states.  This means that ground states of the system correspond to the extreme points in $\Pi$ as we illustrate in Figure~\ref{fig:convexhull} left.  This argument was first given in reference\cite{seko2014efficient} and used to efficiently identify ground states of a cluster expansion.

Consider the limit $\vert v \vert \rightarrow \infty$ such that only the ground state(s) has nonzero probability (as in the previous paragraph). In this case, the model becomes infinitely insensitive to all parameters, i.e., the lowest energy outcomes all become equally probable independent of any variation in the parameters.  This limit therefore corresponds to a vertex (i.e., a zero-dimensional boundary) of the model manifold. 

 In order to find the limits that correspond to higher-dimensional boundaries, we must consider limits in carefully chosen directions.  As $\vert \theta \vert \rightarrow \infty$, it follows from Eq.~\eqref{eq:relatieprob2} that that $P_i / P_j$ remains nonzero and finite only if $\Pi_{i\mu} v_\mu =  \Pi_{j\mu} v_\mu$ which means that the points $\Pi_i$ and $\Pi_j$ must have the same projection onto the vector $v$, i.e., $v$ is perpendicular to the line connecting points $\Pi_i$ and $\Pi_j$, as in Figure~\ref{fig:convexhull}, right.  

We now consider which vectors $v$ reduce the dimensionality of the parameter space by one.  Assume the parameter space has $N$ dimensions and consider the limit $\vert v \vert \rightarrow \infty$.  In this limit, only some of the states will remain with nonzero probability.  As we just argued, the rows of the $\Pi$ matrix corresponding to these non-zero states must lie in a space perpendicular to the vector $v$.  However, in order for this limit to be a manifold boundary of dimension $N-1$, these points must span a space of dimension $N-1$.  Therefore $v$ must perpendicular to an $N-1$ dimensional face of the convex hull of $\Pi$ matrix points.    This is illustrated in Figure~\ref{fig:convexhull} (right) for the case of $N = 2$. 

From this argument, it follows that there is a one-to-one correspondence between $N-1$ dimensional faces of the convex hull of the points in $\Pi$ and the $N-1$ dimensional boundaries of the model manifold.  By repeating this argument for $N-2$, $N-3$, etc. dimensional boundaries, we conclude that the model manifold has the same boundary structure as the convex hull of points in $\Pi$.

\begin{figure}
\includegraphics[width=3.5in]{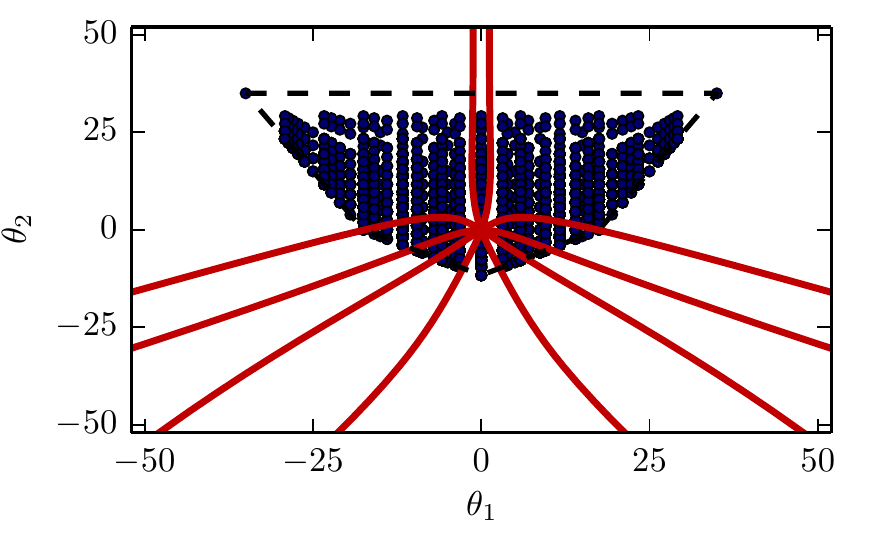}
\caption{\label{fig:fccgeodesics} \textbf{Convex Hull for cluster expansion on an FCC lattice.}  The model manifold is has five edges and five corners (pentagon-like), revealed by the five limiting slopes of the geodesics in parameter space.  This boundary structure is the same as that of the convex hull of the points in $\Pi$ (black dashed line).}
\end{figure}

Geodesics will typically encounter a boundary of dimension $N-1$.  (Only by fine-tuning the initial direction of a geodesic will it encounter a corner of the model manifold.)  This suggests that geodesics in parameter space will asymptotically become perpendicular to the faces of the convex hull of the points in $\Pi$.  We show this explicitly in Figure~\ref{fig:fccgeodesics} for the model whose parameters are the first two non-trivial terms of a cluster expansion of a binary alloy on an FCC lattice.  The first parameter is the on-site energy for having an atom of one type.  The second parameter is the nearest neighbor interaction.  Notice that for large parameters values along the geodesic, the path orients itself to be perpendicular to the faces of the convex hull of points.  In this case, the geodesics paths approach infinity with five distinct slopes, so the model manifold is a pentagon (notice the similarity to Figure~\ref{fig:partition}).

\subsection{Models with Symmetries}
\label{sec:symmetries}

The final general class of models we consider are those with discrete symmetries so that the predictions of the model are invariant under some discrete transformation of the parameters.  The model manifold is therefore isomorphic to a quotient of the parameter space.  Consequently the model manifold will have boundaries associated with the boundaries of the parameter quotient space with boundary points corresponding to fixed points of the parameter transformation.

To make this concept more concrete, consider a model that is invariant to an operation on its parameter space $g$: $P(\theta) = P(g(\theta))$.  Let $\theta^*$ be a fixed pointed of $g$ so that $g(\theta^*) = \theta^*$.  Let $v$ be a vector denoting a direction in parameter space; we are interested in characterizing the circumstances under which the directional derivative of the model in the direction of $v$ will be zero.  If $\partial P/\partial \theta^\mu v^\mu = 0$, then the Fisher Information matrix will be singular in the direction of $v$.  

Let $J = \partial g/\partial \theta$ evaluated at $\theta^*$ and let $v$ be an eigenvector of $J$ with eigenvalue $\lambda \neq 1$.  We can then write
\begin{equation}
  \label{eq:direcderivPG}
  \frac{\partial P}{\partial \theta^\mu} v^\mu \biggr\vert_{\theta = \theta^*}= \lim_{h\rightarrow 0} \frac{P(\theta^* + hv) - P( g(\theta^* + hv) )}{h (1 - \lambda)}.
\end{equation}
It is straightforward to check Eq.~\eqref{eq:direcderivPG} by expanding the terms numerator to lowest order in $h$:
\begin{eqnarray}
  P(\theta^* + hv) & \approx & P(\theta^*) + h \frac{\partial P}{\partial \theta} v \\
  P(g(\theta^* = hv)) & \approx & P\left( g(\theta^*) + h J v \right) \\
  & = & P\left( \theta^* + h \lambda v\right) \\
  & \approx & P(\theta^*) + h \lambda \frac{\partial P}{\partial \theta} v.
\end{eqnarray}

Invoking the invariance of the model under operation $g$ gives $P(\theta^* + hv) - P( g(\theta^* + hv) = 0$.  We therefore conclude that at the fixed point of a symmetry operation, the manifold is non-singular in as many dimensions as $J$ has eigenvalues $\lambda = 1$.

As a concrete example consider the model
\begin{equation}
  \label{eq:2exp}
  y(t, \theta) = e^{-\theta_1 t} + e^{-\theta_2 t}
\end{equation}
with $\theta_\mu \geq 0$.  This model is invariant to the permutation $(\theta_1, \theta_2) \rightarrow (\theta_2, \theta_1)$.  The fixed point of this operation corresponds to the line $\theta_1 = \theta_2$.  This line maps to a boundary on the model manifold as in Figure 1 in reference\cite{transtrum2011geometry} where it is described as a fold line because the model effectively ``folds'' the parameter space along this line.

To see how this geometry relates to the arguments above, consider the Jacobian of the permutation operation:
\begin{equation}
  \label{eq:symmetryjacobian}
  J = \left( \begin{array}{cc} 0 & 1 \\
                       1 & 0 
\end{array} \right)
\end{equation}
which has two eigenvalues, $1$ and $-1$.  The eigenvector with eigenvalue $\lambda = -1$ is $v = (1, -1)^T$ is the parameter space direction perpendicular to the fold line, i.e., perpendicular to the boundary of \MM.  The direction $u = (1,1)$ is a nonsingular direction of the FIM.

Manifolds that locally look like quotient subspaces of $\mathbb{R}^N$ are known as orbifolds.  Depending on the symmetries involved, orbifolds can have a rich boundary structure, a discussion of which is beyond the scope of this paper.  For some orbifolds with more unusual singularity structures (such as a cone), it may not be possible to identify the boundaries using geodesics as we speculate in section \ref{sec:limitations}.  Orbifolds are relevant for this discussion as they are another class of models for which singularities exist that contribute to a hierarchical boundary structure.  Furthermore, there exists a rich mathematical theory to study the global, topological properties and singularity structures of such mappings.

\section{Relation to Other Approximation Methods}
\label{sec:relation}

As we have seen, many boundaries of model manifolds correspond to limiting approximations among its parameters.  These typically correspond to parameters reaching the limit of their physically allowed values.  Approximations of this nature have a long and venerable history in science.  MBAM is an attempt to semi-automate this technique so that it may be applied in new contexts and reveal new insights into complex physical phenomena.  

The converse of this observation is that the countless examples of limiting approximations that have historically been applied to mathematical models can be reinterpreted as special cases of the manifold boundary approximating method.  A continuum limit is a typical example.  A continuum theory, such as a field theory, may be arise when a microscopic length scale, such as a lattice constant, is much smaller than the quantities of interest.  The field theory emerges in the limit that the lattice constant becomes zero, i.e., the boundary of the model manifold.  Similar arguments hold for thermodynamic limits (limits of infinite system size) and various classical limits (limits in which Planck's constant become zero or the speed of light become infinite).  

Although not as obvious, the MBAM is also related to the Renormalization Group (RG).  In particular, note that the Ising model falls into the class of models described in section~\ref{sec:loglinear}, and the Markov random field in section \ref{sec:networks} is an Ising model on a network.  Recall that the Kadanoff block-spin renormalization procedure involves an iterative clustering of spins, not unlike the clustering illustrated in Figure~\ref{fig:mrfexmaple}.  Furthermore, the process of marginalizing a field theory over the highest energy field configurations is analogous to the approximation that those configurations occur with probability zero.  In section~\ref{sec:loglinear}, we showed that models with Hamiltonians linear in the parameters have boundaries that similarly remove a sequence of high-energy configurations from the model.  These superficial connections to RG methods will be explored in more depth elsewhere.

\section{Potential Limitations and Pitfalls}
\label{sec:limitations}

One of the potential limitations to practical application of the MBAM is the computational cost.  Models with many parameters tend to be very complex and computationally expensive.  The most expensive part of the MBAM calculation is repeated calculation of the FIM along the geodesic.  In some models, calculating the derivatives with respect to all of the parameters may be prohibitive.  In other models, it may not be possible to estimate the required expectation values.  Where possible, the calculation of derivatives for the FIM is trivial to parallelize, so the overall calculation scales well with system size.  In our experience, we have successfully applied MBAM to models based on differential equations with hundreds of parameters.  Models of the form of Eq.~\eqref{eq:loglinear} can be explored with even more parameters.

Unlike other, automatic model reduction methods, MBAM is not fully algorithmic.  While most of the steps can be automated, MBAM (at least in its current form) requires human intervention to identify and evaluate the limits in the model.  MBAM is therefore only semi-automatic.  This step is key to MBAM for at least two reasons.  First, it allows MBAM to be very general, effectively adapting itself to the functional form of the model.  Second, it generates theoretical insight into the behavior of the complex model.  The need for human intervention may often be the limiting factor in the size of models tractable by MBAM.  We believe that as the manifold boundaries of specific model classes are better understood, that this problem may be mitigated.  For example, it would be straightforward to fully automate the method for the model class of Eq.~\eqref{eq:loglinear}.  Similarly, we find that in models of chemical kinetics described by differential equations that similar types of limits are repeated so that it may be possible to automate their evaluation and minimize the need for human guidance.  

Calculating a geodesic is incidental to the implementation MBAM.  It is a useful tool for finding boundaries, but for some cases there may be other methods.  For example, for models of the form in Eq.~\eqref{eq:loglinear}, convex hull algorithms applied to $\Pi$ could be used to construct reduced models.  In other cases, it may be possible to theoretically identify relevant boundaries based on symmetries or other arguments.  In some cases, the boundary approximation may be the theoretically desired result although it may be hard to find.

A key to managing the computational cost of the MBAM is identifying a global property of the model manifold (i.e., the boundary) using a sequence of local calculations (i.e., the FIM).  MBAM therefore exploits a nontrivial relationship between global structure and local information that was first identified in reference\cite{transtrum2010nonlinear} (see Figure 3 specifically).  In order for the eigenvalues of the FIM to reflect the model's global structure, the parameters of the model need to be cast in their natural units.  In practice, we do this by transforming to log-parameters.  In some cases, it may be difficult to find a convenient parameterization that makes the FIM eigenvalues useful.  If parameters are poorly scaled, the procedure may encounter difficulties.

MBAM requires that the model manifold have a hierarchy of boundaries.  A major part of this paper was devoted to exploring this structure for a wide variety of models.  Although we have found this structure to be common, one can imagine scenarios that could be problematic for the procedure (i.e., geodesics may not easily identify the desired boundaries) or in which the boundaries do not exist at all.

It is possible that a model may have the desired structure but that it can't be identified by geodesics.  Models with high curvature, for example, may divert the geodesic away from the desired boundary (analogous to a gravitational slingshot).  Another possibility is that the manifold structure may break down at some points.  For example, a cone is a two-dimensional surface that is bounded, not by a one-dimensional line, but by a point.  A singularity of this type could easily arise as the consequence of a symmetry (as in section~\ref{sec:symmetries}).  It may be desirable to approximate such a ``cone'' by its apex, but a geodesic will circle the cone indefinitely without finding the desired point.

In some cases a manifold may be bounded but not have a hierarchy of boundaries.  For example, a model manifold may look more like a sphere than a polyhedron.  In other cases, a manifold may have a ``soft'' boundary, e.g., if one cannot put a hard limit on physical range of a parameter.

Unbounded manifold are also potentially problematic.  We have seen that it is not necessary for the model manifold to be strictly bounded.  The model in Eq.~\eqref{eq:exponentials} is unbounded, but MBAM is still applicable because the cross-sections have the hierarchical structure we seek.  However, if the model has no bounded cross sections, then MBAM will fail.  The simplest example of this is a linear least squares model.  The model manifold is a hyper-plane with no effective low-dimensional structure to approximate.  Another example of an unbounded model manifold is a Markov Random Field (as in section~\ref{sec:networks}) with normally distributed random variables (so that $A$ in Eq.~\eqref{eq:MRFPrecisionMatrix} is the inverse covariance matrix).  

Necessary and sufficient conditions to guarantee the hierarchical boundary structure remains an open problem; a major purpose of this paper is to demonstrate empirically that is shared by a wide variety of models.

\section{Discussion}
\label{sec:Discussion}

In this paper, we have presented a detailed description of the Manifold Boundary Approximation Method (MBAM).  The primary motivation of this model reduction method is to provide a ``gray-box'' representation of a model that is able to bridge the gap between microscopic mechanisms and a system's collective behavior.  It aims to exploit a low-effective dimensionality in the behavior space of an overly parameterized model, using geodesics to find a series of limiting approximations that minimally impact the behavior of the system.  

One of the interesting aspects of the MBAM is its potential to guide theoretical studies of complex systems.  Complex models are often difficult to interpret and ascribing a mechanistic origin to particular behaviors is challenging.  It is often the case that experimental or engineering control knobs operate on a microscopic, mechanistic level of the system.  In these cases (including much of material science, biology, neuroscience, etc), a way to effectively connect the complicated mechanistic description to a simple phenomenological description would be useful.  

In these cases, effective model reduction does not simply fix a technical problem that arises from not having a big enough computer.  Rather, it is an ongoing process that provides theoretical insight and refines one's view of the system.  The whole process ought to provide feedback to the original, complicated model that incorporates the insights gained from the simple representation.  Because of its ``gray-box'' structure, MBAM is an effective complement to other automatic model reduction methods that produce ``black-box'' approximations in a way that accommodates this sort of ongoing refinement.

One of the curious requirements of MBAM is that the model manifold have a hierarchical structure of boundaries.  One of the primary purposes of this paper is to demonstrate that this structure is common to many models.  Although there are counter examples (as in section \ref{sec:limitations}), it is surprising that this structure is so common.  The reason for the ubiquity of this structure remains an open question.  

It is also interesting to note that many common methods for constructing effective mechanisms (such as continuum limits, singular perturbation, steady-state and partial equilibrium approximations) implicitly use this hierarchical boundary structure.  By identifying these diverse approximations as a common geometric operation, MBAM is a step toward unifying and automating this process.  Indeed, one of the challenges to constructing reduced representations in complex systems is that one cannot easily identify a priori which approximations, among the many choices, one should make.  Constructing approximate models therefore requires expert guidance built on years or decades of hard-won intuition.  MBAM is a promising tool for identifying these approximations automatically and in a reproducible, mathematically rigorous way. Application to complex models may reveal previously unknown classes of approximations.  Our hope is that MBAM may motivate useful approximations that in turn reveal which mechanisms govern complex nonlinear phenomena.

\textbf{Acknowledgments:}  The author thanks Peng Qiu for helpful discussions and Gus Hart for providing the $\Pi$ matrix in Figure~\ref{fig:fccgeodesics}.

\bibliography{../../References/References}

\end{document}